\def\bfig{\begin{figure}}  
\def\efig{\end{figure}}

\def\gh{{\bf \hat g}}

\def\inout{{\stackrel{\rm in}{\rm out}}}

\def\beq{\begin{equation}}
\def\eeq{\end{equation}}

\def\beqa{\begin{eqnarray}}
\def\eeqa{\end{eqnarray}}

\documentstyle[a4]{article}

\let\ssection=\section
\renewcommand{\section}{\setcounter{equation}{0}\ssection}

\begin{document}
\setlength{\unitlength}{0.7cm}
\openup 1.5\jot

\begin{titlepage}
\hfill SWAT/94-95/57
\hfill UM-P-94/122
\hfill QMW-PH-94-39
\hfill hep-th/9501078
\vskip 2cm
\begin{center}
\huge

Restricted Quantum Theory of Affine Toda Solitons

\vskip2cm
\Large
B. Spence
\vskip 1cm\large
{\it School of Physics\\
 University of Melbourne\\
 Parkville 3052 Australia}
\footnote{Current Address: Physics Department, Queen Mary \& Westfield
              College, Mile End Road London E1 4NS UK }\\
{\tt E-mail:B.Spence@qmw.ac.uk}
\vskip 2cm
\Large
Jonathan Underwood
\vskip 1cm\large
{\it Physics Department\\
University College of Swansea\\
Swansea SA8 2PP UK}\\
{\tt E-mail:J.Underwood@swan.ac.uk}
\normalsize
\vskip 2cm
\end{center}
\begin{abstract}

We quantise the reduced theory obtained by substituting the soliton
solutions of affine Toda theory into its symplectic form. The
semi-classical S-matrix
is found to involve the
classical Euler dilogarithm function.

\end{abstract}
\end{titlepage}

\section{Introduction}\label{intro}

The affine Toda field theories associated to an affine Kac-Moody
algebra $\gh$ have received a great deal of attention in recent years.
The classical equation of motion which describes them is

\beq \Box^2
\phi+{2\mu^2\over\beta}\sum_{i=0}^r m_i H_i
e^{\beta\alpha_i(\phi)}=0,\label{affext2}\eeq
where $\phi$ is the dynamical field, taking values in the Cartan
subalgebra of $\gh$ (for whom Chevalley generators are the $H_i$ and
the Coxeter number is $h$), $\mu$
and $\beta$ are real parameters, and $\alpha_i$ are the simple roots
of $\gh$. Taking $\beta$ to be real we obtain a model with a particle
spectrum whose properties are algebraic in origin (see, for example,
refs.
\cite{FLO91,Do92}), and whose S-matrix is exactly known
\cite{Do91,Do92,BCDS89a,BCDS89b,BCDS90,FO92,KM90,CD91,MF91}.

If the coupling constant $\beta$ is pure imaginary then the model
possesses a series of vacuum solutions. We expect soliton
solutions to be those of minimal energy which interpolate these vacua,
the difference between the final and the initial vacua being known as
the topological charge of the soliton. The seminal calculation was
made by Hollowood  \cite{Ho92} for the $\hat A_r$ models using
Hirota's method. This technique was later developed to include the
rest of the theories \cite{ACFGZ92,CZ93}. A new method,
involving the use of the general solution discovered by Leznov and
Saveliev (see e.g. \cite{LS92}), has recently proved to be very useful in
extracting particular properties of the solitons \cite{OTU93}. This method can
be
motivated by consideration of the form of the canonical
energy-momentum tensor
\beq
T_{\mu\nu}= \left(\partial_\mu\phi ,\partial_\nu \phi\right) -
{g_{\mu\nu}\over 2}
\left(\partial_\alpha \phi,\partial^\alpha
\phi\right)
-\frac{2g_{\mu\nu}}{\beta^2}
\sum_{k=0}^{r}n_k\left(e^{\beta{\alpha}_k(\phi)}-1\right).\label{em}\eeq
Using arguments given in ref. \cite{OTU93} it can be shown that
$T_{\mu\nu}$ can be split into a sum of two parts,
$T_{\mu\nu}=C_{\mu\nu}+\Theta_{\mu\nu}$, where $\Theta_{\mu\nu}$ is traceless
and $C_{\mu\nu}$ is a total derivative.
Due to the topological nature of solitons, their properties, in particular
their energies and momenta,
depend only on their behaviour at
infinity, and hence we might expect $\Theta_{\mu\nu}$ to vanish for soliton
solutions.
Further work \cite{OTU93,Un93} shows that $\Theta_{\mu\nu}$ can be written
purely in terms of the chiral fields which are the parameters of the
Leznov-Saveliev solution, so it is plausible that the soliton solutions
arise if these fields vanish.

In order to find the N-soliton solutions a further ansatz for the form
of a constant of integration appearing in the general solution must be
made \cite{OU93}; the N-soliton specialised solution turns out to be
\beq e^{-\beta\phi_k} =\langle\Lambda_k
|\exp\left(W_1\hat F^{i_1}(z_1)\right)\ldots
\exp\left(W_N\hat F^{i_N}(z_N)\right) |\Lambda_k\rangle. \label{fulsol} \eeq
For a detailed explanation of the quantities appearing in the formula
the reader should consult ref. \cite{OU93}; we will need only the following
details. The parameters $i_j$ are integers lying between 1 and $r$
which describe the species of the N solitons. $z_j$ are complex
parameters related to the soliton rapidities $\rho_j$ in the following way:
\beq z_j=\theta_j
e^{-\rho_j}\frac{|q^+_{i_j}|}{q^+_{i_j}},\label{onereal}\eeq
where $\theta_j=\pm
1$, and the $q_{i_j}$ are structure constants of $\gh$ which can be calculated
from the following formula of ref. \cite{FLO91}
\beq q_p^+\equiv \gamma_p\cdot
q(1)=2\imath x_p(1)e^{-\delta_{pB}\imath\nu\pi/h} \label{gq} \eeq
($\i \equiv \sqrt{-1}$). The functions $W_j$
encode the space-time dependence of the solution,
and are given by the formula
\beq W_j= Q_j\exp \left\lbrace \sqrt 2\mu\theta_j\mid
q_{i_j}^+\mid
\left(t\sinh \rho_j- x\cosh\rho_j\right)\right\rbrace.\label{wndef}\eeq

The first major use to which this formalism was put was a calculation
of the masses of single solitons of species $p$. These turned out to
be \cite{OTU93}:
\beq M_p=-{4\sqrt
2h\over\beta^2\gamma_p^2}\mu |q^+_{p}|.\label{mass}\eeq
An interesting fact which is not yet fully understood is that these
masses are in a certain sense dual to those of the particles.

In this letter we will use further features of the solution
(\ref{fulsol}) to calculate the Poisson brackets on the N-soliton phase
space and will perform a canonical
quantisation to extract the S-matrix.

\section{Poisson Brackets}\label{pobr}

The classical phase space of the affine Toda theories has been discussed
recently in ref. \cite{PS94}.
 The
symplectic form $\Omega$ is calculated by integrating the symplectic current
over all space at some time,
and is simply
\beq
\Omega=\int dx \left(\delta\phi,\delta\partial_t\phi\right)\label{sform}\eeq
where $(\ ,\ )$ denotes the usual Killing form on the algebra $\gh$.

We wish to investigate the phase space of the soliton solutions. To do this,
we insert the soliton solutions
into the symplectic form $\Omega$.
First let us calculate the symplectic form for a one-soliton solution.
We will need to use two important properties of this solution, but our
method will mean that we do not need the explicit form of the solution
itself. This is a typical feature of calculations concerning
solitons in affine Toda theory. The first feature we need is that the soliton
is a relativistic object, and so the Poincare algebra must be
realised on the phase space. What this means for the solution is that
the field $\phi$ must appear only as a function of $u$, where
\beq u=t\sinh\rho
-(x-x_0)\cosh\rho.\label{udef}\eeq
The rapidity $\rho$ and centre of mass $x_0$ are the real parameters of the
soliton solution. The relationship of these parameters to those
in the algebraic ansatz for the soliton solution, as well as the explicit
dependence of $\phi$ upon $u$, can be determined from the formul\ae\ of
\cite{OTU93}, \cite{OU93}, but we shall not need these yet. Using the
antisymmetry properties of the wedge product we obtain
\beqa \Omega &=& \int dx
\left(\frac{d\phi}{du},\frac{d\phi}{du}\right)\; \cosh\rho\, \delta x_0\wedge
\delta
(\sinh \rho), \nonumber \\ &=& -\int dx
\left(\partial_t\phi,\partial_x\phi\right)\; \frac{\delta x_0\wedge
\delta
(\sinh \rho)}{\sinh\rho},\nonumber \\&=& -\int
T_{tx}\; \frac{\delta x_0\wedge \delta (\sinh \rho)}{\sinh\rho},
\label{step}\eeqa
where the last step is performed using the
expression for the energy-momentum tensor (\ref{em}).
Finally,
substituting $P=-\int dx T_{01}= M\sinh\rho$ into (\ref{step}),  we obtain
\beq \Omega=\delta \xi\wedge\delta \rho,\label{onesym}\eeq
where $\xi=Mx_0
\cosh\rho$ is the canonical variable conjugate to $\rho$. This is of
course the result we expect.

In order to evaluate the symplectic form in the more general case of
$N$ solitons we consider the form of the solution as $t\rightarrow\pm \infty$.
In
the generic situation the solitons will be
well separated and since $\Omega$ is a
local expression we can just add up the contributions from each of the
solitons in turn. Thus we find
\beq \Omega= \sum_{i=1}^N \delta \xi_i^\inout\wedge
\delta \rho_i^\inout.\label{inout}\eeq
We now need
to relate these variables to those which parameterise the solution.

Let us consider the form of the solution (\ref{fulsol}) as $x$ increases
from $-\infty$.
The first significant departure from the vacuum will
occur with the soliton of greatest rapidity, $\rho_N$. We know from
ref. \cite{OU93} that the greatest non-vanishing power of $\hat F^{i_N}$
within a representation of level $x$ is $x$.\footnote{We are only
considering the theories where $\gh$ is simply-laced from now on.}
This in turn means that the solution for the component field $\phi_k$
will be the logarithm of a polynomial of degree $m_k$ in $W_N$ defined
by equation (\ref{wndef}).
As we move through this soliton $W_N$ becomes much greater than 1 and
so we can ignore all but the highest power in the polynomial as far as
calculating the form of the solution for greater $x$ is concerned.
This term of course multiplies an algebraic factor $\hat
F^{i_N}(z_N)$. Normal  ordering  the vertex operator expression
\cite{KO93}, \cite{OU93} for this yields
\beqa \lefteqn{e^{-\beta\phi_k} =\langle\Lambda_k
|F^{i_N}(z_N) |\Lambda_k\rangle W_N^{m_k}\langle\Lambda_k
|\exp\left(X_{i_1,i_N}(z_1,z_N)W_1\hat F^{i_1}(z_1)\right)
\ldots}
\label{nextsol}
\\ &&\hspace{2cm}\ldots
\exp\left(X_{N-1,N}(z_{N-1},z_N) W_{N-1}\hat
F^{i_{N-1}}(z_{N-1})\right) |\Lambda_k\rangle, \nonumber \eeqa where
\beq X_{i,j}(z_1,z_2)=\prod_{n=0}^{h-1} \left(1-e^{-2\pi
in/h}\frac{z_2}{z_1}
\right)^{w^n(\gamma_i)\cdot\gamma_j}.\label{psmat}\eeq
The roots
$\gamma_i$ are the simple roots $\alpha_i$ up to a sign \cite{OU93},
and $w$ is the Coxeter element of the Weyl group.
Thus, aside from a constant shift in the field $\phi$, the only effect
is to change each of the $Q_j$ for $j<N$ by a factor
$X_{i_j,i_N}(z_j,z_N)$.  Extending this argument as we move through each
of the solitons in turn leads us to conclude that as $t\rightarrow
-\infty$ the field $\phi$ becomes a sum of one-soliton solutions but
with new parameters
\beq Q_j^{\rm in}=Q_j\prod_{p>j}X_{j,p}(z_j,z_p).\label{qin}\eeq
The rapidities remain unchanged under the transformation to `in'
variables. Now all we need to do is relate the variables $Q_j$ to the
$\xi_j$. Comparing expressions (\ref{wndef}) and (\ref{udef}) we find that
up to an irrelevant constant
\beq x_0= \frac{\theta\ln Q}{\sqrt 2\mu|q^+_i|\cosh\rho}.\label{bored}\eeq
Using
the mass formula equation (\ref{mass}) we obtain
\beq\xi=\frac{2h\theta\ln Q}{|\beta|^2}\label{xicq},\eeq
and so
\beqa \xi_j^{\rm in} &=& \xi_j + \frac{2h}{|\beta|^2} \sum_{p>j}\ln
X_{j,p}(z_j,z_p),\nonumber \\ \rho^{\rm
in}_j&=&\rho_j.\label{inpar}\eeqa
Similar arguments yield the following relationships
for the out variables
\beqa \xi_j^{\rm out} &=& \xi_j + \frac{2h}{|\beta|^2} \sum_{p<j}\ln
X_{p,j}(z_p,z_j),\nonumber \\ \rho^{\rm
out}_j&=&\rho_j.\label{outpar}\eeqa
Since the symplectic form $\Omega$
is trivial to invert in terms of terms of either `in' or `out'
variables the appropriate Poisson brackets for the natural variables
follow straightforwardly, and can be seen to be consistent if we note
that $X(z_j,z_p)$ depends only on the rapidity difference
$\rho_j-\rho_p$.

\section{Scattering Matrix}\label{scat}

Having found canonical coordinates on the classical phase space we can
now quantise the theory in a straightforward manner, simply by
replacing the canonical Poisson bracket with the canonical commutators
\beq \left[\xi^{\rm in},\rho^{\rm in}\right]=
\left[\xi^{\rm out},\rho^{\rm out}\right]
       = \imath\hbar.\label{cancom}\eeq
We can then attempt to discover the unitary transformation (analogue
of the canonical transformation) which will be the S-matrix for the
reduced theory. From equations (\ref{inpar}) and (\ref{outpar}) we
conclude that
\beqa \xi^{\rm out}_j &=& \xi^{\rm in}_j
+\frac{2h}{|\beta|^2}\sum_{p< j}\left(\ln
X_{p,j}(z_p,z_j)-\sum_{p> j}\ln
X_{j,p}(z_j,z_p)\right)\nonumber \\ \rho^{\rm out}_j&=&\rho^{\rm in}_j
=\rho_j. \label{inout2}\eeqa
We can see from these equations that the S matrix
$S$ is purely a function of the rapidity differences, and satisfies the
equation
\beq {\partial ({\rm log} S)\over \partial\rho_j} = {-2h{\it\i}\over\vert
\beta\vert^2\hbar}
   \left(\sum_{p<j} {\rm ln} X_{p,j}(\rho_p,\rho_j) -
\sum_{p>j} {\rm ln} X_{j,p}(\rho_j,\rho_p) \right).
\label{anotherone}\eeq
The solution of this is
\beq S=\exp\left\lbrace
\frac{2h\imath}{|\beta|^2\hbar}\sum_{b> a}
\sum_{n=0}^{h-1}w^n(\gamma_a)
\cdot(\gamma_b)\; Li_2\left[\exp\left(
\rho_a-\rho_b+\frac{\imath\pi}{h}\left(\delta_{i_b,B}-\delta_{i_a,B}-2n
\right)\right) \right]
\right\rbrace
.\label{shakeitallabout}\eeq
We have used the definition of the classical Euler dilogarithm
\beq Li_2(v)=\sum_{s=1}^{\infty}\frac{v^s}{s^2}=-\int_0^v
{\ln\left(
1-y\right)\over y} dy.\label{whatacorker}\eeq
Many of the interesting properties
of this function were studied by one William Spence
in the early nineteenth century \cite{S26}. A more recent discussion
in the mathematical literature can be found in ref. \cite{Le81}.
Recent developments relating to conformal field theory are reviewed
in ref. \cite{K94}.
The matrix (\ref{shakeitallabout}) satisfies various conditions, which
are essentially related to properties of the time-delay functions $X_{ij}$.
This function was studied recently in ref. \cite{FJKO94}, where it was noted
that it has properties
reminiscent of those of $S$ matrices. Indeed, one can think
of these properties as being consequences of the fact that the $S$ matrix
(\ref{shakeitallabout})  
is periodic, symmetric, etc. Let us define
\beq  T_{ab}(\rho) =
\sum_{n=0}^{h-1}w^n(\gamma_a)
\cdot(\gamma_b)\; Li_2\left[\exp\left(
\rho + \frac{\imath\pi}{h}\left({c(a) - c(b) \over 2} - 2n
\right)\right) \right].
\label{why bother} \eeq
Then the matrix function $T_{ab}(\rho)$ satisfies the following relations:
\begin{description}
\item[{\rm(i)}] $T_{ab}(\rho+2\i\pi) = T_{ab}(\rho)$
\item[{\rm(ii)}] $T_{ab}(\rho) = T_{ba}(\rho)$
\item[{\rm(iii)}] $T_{ab}(\rho+i\pi) = -T_{\bar a b}(\rho)$
\item[{\rm (iv)}] $(T_{ab}(\rho^*))^* = T_{ab}(\rho)$
\item[{\rm (v)}] $T_{ab}(\rho) = -T_{ab}(-\rho) + 2T_{ab}(0)$
\item[{\rm (vi)}] $\sum_{t=i,j,k} T_{lt}(\rho+\i\eta_t) = 0$, where $l$ is
a free label and $i,j,k$ satisfy a fusing rule; $\eta_t=-2\xi_t +
{c(t)-1\over2}$,
where $c(t)$ is $1(-1)$ if the root $\alpha_t$ is black(white), and $\xi_t$
are the integers in the fusing relation
$\sum_{t=i,j,k}\omega^{-\xi_t}\gamma_t=0$.
\end{description}
These relations can be shown directly:
Property (i) is obvious; properties (ii), (iii) and (iv)
follow using arguments analogous to those
used for the time-delay functions
$X_{ab}(\rho)$ in ref. \cite{FJKO94}.
Property (vi) similarly follows from an argument analogous to that given in
ref. \cite{FO92} when discussing the affine Toda particle $S$-matrix.
Property (v) is most easily proved by noting firstly that the
$\rho$ derivative of this equation is true - this
follows using eqn. (\ref{whatacorker})
and the fact that $X_{ab}(\rho)=X_{ab}(-\rho)$
(see ref. \cite{FJKO94}). Thus the left-hand side of (v) equals the
right-hand side up to an additive constant, and
putting $\rho=0$ one sees that this constant must
vanish. A direct proof of (v) seems more involved - for example, for the $A_1$
case
this property reduces to the equation
\beq Li_2(-e^\rho) - Li_2(e^\rho) = -Li_2(-e^{-\rho}) + Li_2(e^{-\rho})
                    +2Li_2(-1) - 2Li_2(1). \label{brmmm}\eeq
The Euler dilogarithm satisfies the following \lq inversion' relation
\cite{K94}
\beq Li_2(-y) + Li_2(-1/y) = -{\pi^2\over6} - {1\over2}({\rm log}\, y)^2.
\eeq
The function $Li_2(y)$ is divergent for
real $y$, $y>1$; however, one can define a continuous function on the real line
by setting \cite{K94}
\beq
     Li_2(y) = {\pi^2\over3} - Li_2(1/y) - {1\over2}({\rm log}\,y)^2, \qquad
    {\rm for}\;\; y>1.
\eeq
Then, using the above two equations and the facts that $Li_2(1)=\pi^2/6$ and
$Li_2(-1)=-\pi^2/12$, equation (\ref{brmmm}) may be proved.

The constant term in the relation (v) means that the S matrix
(\ref{shakeitallabout})
will be invariant under $\rho\rightarrow-\rho$ only if it is normalised so that
$S(\rho=0)=1$. Physically this requirement is obvious - that there be no
scattering
when the two solitons have the same rapidity.
We note that the properties (i)-(vi) above reflect identities satisfied by
the Euler dilogarithm.

Thus the expression (\ref{shakeitallabout}) satisfies relations expected for
an $S$-matrix. Note, however, that the Euler dilogarithm can be continued
to a multi-valued function on the complex plane, minus the segment
$(1,\infty)$ of the real axis \cite{Le81,K94}.
Hence our proposed $S$-matrix does not have the expected pole structure.
We will comment upon this in the following section.

\section{Conclusions and Developments}\label{theend}

It is important to be clear about what this S-matrix is, and what it
is not. We have only made a semi-classical approximation to the
quantum theory of affine Toda solitons, and so do not expect to see
all of the behaviour typical of a quantum field theory. In particular
as noted above, there are no poles in formula (\ref{shakeitallabout}), and we
have made no mention of the renormalisation of the soliton masses. The
interested reader should consult the recent papers \cite{MW94} and
\cite{DG94} for a discussion of this latter point. Very little is
known about the expected behaviour of the poles on account of
difficulties concerning the unitarity of the affine Toda theories in
the imaginary coupling r\'egime. Another feature expected of the
quantum theory which this kind of approximation will not possess is
that there will be no processes which change the topological charges
of the scattering solitons. This is because such these are absent
classically. An obvious conjecture worthy of exploration is that
a full quantum version of (\ref{shakeitallabout}) involves the
{\it quantum} dilogarithm
of ref. \cite{FK93}.

What we believe we have is the S-matrix of an quantum integrable
particle theory, generalising that of Ruijsenaars \cite{Ru,Ru2}. In
the case of $\hat A$-series solitons of equal mass this particle model
already reproduces the appropriate scattering shifts. In the more
general case however, no such model is known. Starting from the formula for
the shifts which follows from Ruijsenaars model, and fixing the
rapidities in such a way that the soliton fusing rule \cite{OU93} is
satisfied, we have obtained correct expressions for the shifts
when two solitons of different masses scatter (it is reasonably easy
to show that all of the $\hat A$-series solitons can be obtained by
repeated fusing from the lightest in this manner). The question is
whether this procedure can be implemented at the level of the
Ruijsenaars Hamiltonian. This is a non-trivial problem since the
fusings correspond to imposing imaginary constraints on the particle
rapidities, and it is unclear what effect this will have on the
symplectic structure of the phase space. We remark that the change
$\rho\rightarrow \rho+ 2in\pi/h$ is a symplectic transformation,
albeit a complex one, which gives us some hope that the above
procedure can be made to work.

A natural generalisation of this work would seem to be to try and
include the breathers. In the sine-Gordon theory at least these
breathers are related to the particles, and possess a number of
discrete energy levels according to the value of the coupling constant
$\beta$. It is a long-standing problem in affine Toda soliton theory
to try and characterise the breather spectrum for more general
theories, and in particular to see if the particle-soliton
correspondence remains. We have obtained preliminary results in
this direction, including a description of the energy levels of the
$\hat A$-series breathers of remarkable and rather mysterious
simplicity. We plan to discuss these issues in a forthcoming paper \cite{SU95}.
\hfill\break

\noindent{\bf Acknowledgements}\hfill\break
\noindent Part of this work was done under the auspices of a Royal Society
Visiting Fellowship award to one of us (JWRU). JWRU thanks the
late UK Science and Engineering Research Council, and
BS acknowledges
 support from the Australian Research Council and
the UK Engineering and Physical Sciences Research Council.
\bibliographystyle{unsrt}

\end{document}